\documentclass[aps,prl,preprint,groupedaddress]{revtex4-1}
\usepackage{amssymb}
\usepackage{amsmath}
\usepackage[figuresright]{rotating}
\begin{document}

\title{Isolated Structures in Two-Dimensional Optical Superlattice}

\author{Xinhao Zou, Baoguo Yang, Xia Xu, Pengju Tang and Xiaoji Zhou$^{1,*}$}

\affiliation{}

\date{\today}

\begin{abstract}
Overlaying commensurate optical lattices with various configurations
called superlattices can lead to exotic lattice topologies and, in turn,
a discovery of novel physics. In this study, by overlapping the maxima of
lattices, a new isolated structure is created, while the
interference of minima can generate various ``sublattice''
patterns. Three different kinds of primitive lattices are used to demonstrate isolated square, triangular, and hexagonal ``sublattice''
structures in a two-dimensional optical superlattice, the patterns of which can be manipulated dynamically by tuning the polarization, frequency, and intensity of laser beams. In addition, we propose the method of altering the relative phase to adjust the tunneling amplitudes in ``sublattices.''
Our configurations provide unique opportunities to study particle entanglement in ``lattices'' formed by intersecting wells and to implement special quantum logic gates in exotic lattice geometries.
\end{abstract}


\maketitle

\section{Introduction}
The system of ultracold atoms trapped in optical lattices is a versatile tool to
simulate systems in condensed matter physics and quantum information
processing \cite{intro1,intro2,intro3,intro4}. The lattice is generated by the
atom-photon interaction and the setup can be one-, two-, or three-dimensional. While primitive Bravais lattices are
constructed in most experiments with a single site per unit
cell \cite{introSFMI,introtriangular}, there has been a growing
theoretical and experimental effort to build non-standard optical
lattices with a few-site basis called superlattices and study the
quantum information processing therein \cite{introdoublewell,introkagome,introkagome1}.
One- and two-qubit gates have been realized in a standard two-dimensional (2D) superlattice \cite{quanOneD,quanlogic,squre}.
The potential application is so diverse that a search for more
exotic superlattices and a higher degree of tunability in the lattice
parameters have been of significant interest.

In contrast to the coupling strengths in conventional 2D superlattices, those of two neighboring particles are the same along all directions and the isolated structures have different interplaquette coupling strengths; therefore, we can study the interaction between two singlets \cite{squre} in isolated square superlattices. Furthermore, since the triangular and hexagonal superlattices are framed by three laser beams, there are three equivalent directions in such structures \cite{isolateTriangular1,isolateTriangular2,isolateTriangular3,isolateTriangular4,isolateTriangular5}. Therefore, we can find the entanglement of three particles or three dimers, which is similar to the Efimov state \cite{efimov}.

In this paper, we propose several schemes to realize isolated
structures in a 2D optical superlattice. The lattice potentials are
calculated using the $E^{[2+\epsilon]}$ method \cite{magic,Esecondorder}, which takes atom-photon interactions as well as hyperfine interactions into
account and is therefore appropriate for computing the light shift of
hyperfine energy levels. By overlapping the maxima of lattices,
the isolated structures are created, while the interference of
minima can generate various ``sublattice'' patterns. We can also
control the superlattice parameters such as symmetries and barrier
height of ``sublattices'' by simply tuning the relative phase of the laser beam.
In a 2D optical lattice, structures such as hexagonal,
triangular, and square have been realized experimentally; hence, we overlay
two such structures with different periods and construct the
isolated hexagonal lattice, isolated triangular lattice, and 
isolated square lattice by using the superlattice techniques.

The remainder of this manuscript is organized as follows. In Sec. II,
we briefly introduce the $E^{[2+\epsilon]}$ method for calculating
the ac Stark shift of hyperfine levels of alkali-metal atoms. In
Sec. III, we show how to realize various isolated lattice structures and how the geometry of the sublattice can be tuned
dynamically. In Sec. IV, we discuss the
potential application of our configurations.

\section{The $E^{[2+\epsilon]}$ method for optical lattice potentials}
An optical lattice is a light-induced periodic potential in
space. In order to calculate the potential ``felt'' by an atom in
its hyperfine states, we have to treat both the atom-photon
interaction and hyperfine interaction as perturbations. Here,
we briefly introduce the $E^{[2+\epsilon]}$
method, which can be used to calculate the lattice potential
for hyperfine levels of alkali-metal atoms.

In the dipole approximation, the Hamiltonian of atom-photon
interaction is ${H}_e=\vec{p}\cdot \vec{E}$, where $\vec{p}=-e\vec{r}$ is the
electric dipole moment, $\vec{r}$ the position of the
electron, $e$ the elementary charge, and $\vec{E}$ the electric field vector. Using the Born-Oppenheimer approximation and the long-wavelength approximation,
this expression can be further simplified as
\begin{eqnarray}
\begin{split}
\vec{E}(\vec{R},t)&=K(\vec{R})e^{-i\omega_Lt}+K^\dagger(\vec{R}) e^{-i\omega_Lt},\\
K(\vec{R})&=\sum_{i}\hat{\epsilon_i}e^{i\vec{k_i}\cdot\vec{R}},\\
\end{split}
\label{hamiltonian}
\end{eqnarray}
where $\vec{R}$ is the position of the nuclei. Here, we only
consider the far-off-resonance laser fields with the same angular frequency $\omega_L/2$ but with different wave vectors $\vec{k_i}$
and polarizations $\hat{\epsilon_i}$. The total light shift of an
atom is simply the sum of light shifts induced by laser fields with
different frequencies.

\begin{figure*}[t]
	\begin{center}
		\includegraphics [width=14cm,height=8cm]{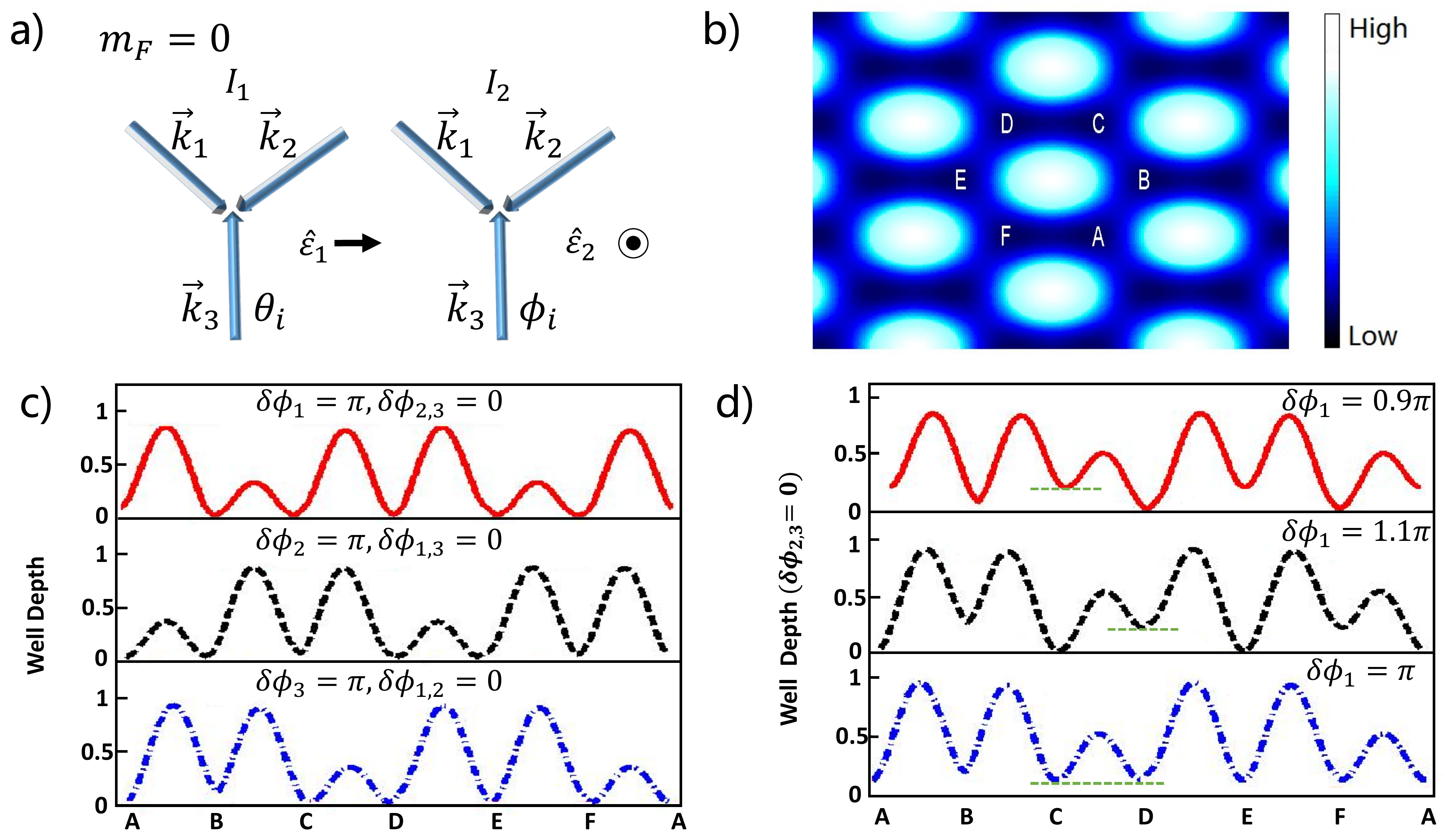}
	\end{center}
	\caption{ (Color online) The double well configuration in a hexagonal
		lattice. (a) The setup for $m_F=0$ state. The lattice is formed by
		two sets of three-laser-beam systems with in-plane and out-of-plane
		polarizations. (b) The 2D optical potential for a
		left-right double well configuration is shown with a scaled
		colormap. A unit cell contains six sites labeled $A, B, C, D, E$ and
		$F$. The double well is formed between two potential minimums such
		as $F-A$ and $D-C$. (c) Optical potential along the channel
		$A-B-...-F-A$. Adjusting the relative phases and making one
		$\delta\phi_i=\theta_i-\phi_i=\pi$, $i=1,2,3$, meanwhile keeping the
		other two $\delta\phi_i=0$ allows a site pairing with all three
		nearest neighbors, therefore the left-right (blue dot-dashed),
		lower-left (black dashed) and lower-right (red solid) double well
		configurations can be created, respectively. A double well is formed
		by two sites with a lower barrier between them. (d) Optical potential
		along the channel $A-B-...-F-A$. Small deviation of $\delta\phi_i$
		from $\pi$ can break the double well symmetry and the relative
		offset between adjacent sites can be tuned. A demonstration using
		left-right double well configuration is shown where the symmetry in
		$F-A$ or $D-C$ is broken. The value of $\delta\phi_{1}$ deviate from
		$\pi$ while $\delta\phi_{2,3}=0$.} \label{rep_diagram1}
\end{figure*}

We now calculate the lattice potential for the energy level
$|i\rangle=|n_i I J_i F_iM_{F_i}\rangle$, where $n$ is the principal
quantum number, $I$ is the nuclear spin, $J$ is the electronic total
angular momentum, and $F=I+J$ is the total angular momentum. $M_I$,
$M_J$, and $M_F$ are the projections of $I$, $J$, and $F$ on the $z$
axis, respectively, where the $z$ axis is taken as the quantization
axis. For hyperfine structures, both the hyperfine interaction and
the atom-photon interaction should be treated as perturbations, and
the optical lattice potential $U_{i}(\vec{R})$ is
\begin{eqnarray}
\begin{split}
&U_{i}(\vec{R})=
\\&-{3 \pi c^2 I_{L}}\sum\limits_{j\neq
	i}\frac{A_{\rm{J_{ji}}}(2F_j+1)(2F_i+1)(2J_j+1)\omega_{\rm{F_{ji}}}}{\omega^3_{\rm{J_{ji}}}(\omega^2_{\rm{F_{ji}}}
	-\omega^2)}
\\& \times\left\{\begin{array}{ccc} J_i& J_j& 1\\F_j&F_i
&I\end{array}\right\}^2\sum_{p=0,\pm 1}{\left(\begin{array}{ccc}
	F_{j}& 1 & F_i\\M_{\rm{F_j}}
	&p&-m_{\rm{F_i}}\end{array}\right)^2|K_p(\vec{R})|^2},
\end{split}
\label{lightshift1}
\end{eqnarray}
where $I_L=(\hbar N \omega)/(\epsilon_0V)$ is the light intensity,
$A_{\rm{J_{ji}}}$ is the Einstein coefficient for the fine-structure
transition between $|i\rangle$ and $|j\rangle$,
$\hbar\omega_{J_{ji}}$ is the difference between fine-structure
energies, and $\hbar\omega_{F_{ji}}$ is $\hbar\omega_{J_{ji}}$ plus
the difference between hyperfine energy corrections. The term inside
curly brackets is a $6J$ symbol, while that inside the large round brackets
is the $3J$ symbol, which describes the selection rules and relative
strength of the transitions. The only term concerning the light
fields is $K_p(\vec{R})$, which is the pth ($p=0,\pm 1$) spherical
component of the rank-1 tensor $K(\vec{R})$. From another perspective,
it also stands for the different polarization components of the
laser fields ($p=1,0,-1$ for right-handed, linear, and left-handed
polarization, respectively). As we can see, $E^{[2+\epsilon]}$
has a very compact form and can easily incorporate experimental
atomic structure data; therefore, it is useful for the lattice-potential
calculations of many elements.

Here, we take the hyperfine ground states $|F=1,m_F=0,\pm1\rangle$ of
$^{87}Rb$ as an example and further simplify Eq.
(\ref{lightshift1}). Because $5s^2S_{1/2}-5p^2P_{1/2}$ and
$5s^2S_{1/2}-5p^2P_{3/2}$ are the two primary transition lines with
a much larger Einstein coefficient $A_J$ and much smaller transition
energy $\omega_J$, we can neglect other excited energy levels and
only consider these two lines as a good approximation. In the
following sections, a beam is regarded as red or blue detuned only
when it is red or blue detuned to both the transition lines. The
formula Eq. (\ref{lightshift1}) includes the scalar, vector, and
tensor light shifts.

\section{Adjusting superlattice parameters for novel lattice geometries}

To construct the isolated structures, we need to not only overlay
commensurate lattices with different periodicity, but also tune
their relative positions and depth, which requires a high degree of
tunability in experimental setups. One advantage of the optical lattice
is that the lattice parameters can be controlled by adjusting the
intensity, polarization, and relative phase of the laser beam,
and an even richer variety of lattice structures can be constructed
and controlled if we operate in superlattices. Here, we take double
wells in a 2D hexagonal lattice as an example and
demonstrate the new lattice geometries created in these
superlattices.

A hexagonal optical lattice can be constructed using three laser
beams with equal intensity and equal frequency that intersect in the xy
plane with mutually enclosing angles of $2\pi/3$ \cite{introtriangular}.
The lattice potential is modulated for different Zeeman
levels \cite{zeeman}. As shown in Fig. (\ref{rep_diagram1}a), the
double-well configuration can be realized for the $m_F=0$ state by using
two sets of three-laser-beam systems: one with wave vectors
$\vec{k_i}$ and in-plane polarizations $\hat{\epsilon_1}$, and
the other with the same wave vectors but out-of-plane polarizations
$\hat{\epsilon_2}$. The in-plane polarized beams have intensity
$I_1$ and phases $\theta_i$, while the out-of-plane
polarized beams have intensity $I_2$ and phases $\phi_i$.
Each set will generate a hexagonal lattice, and changing the phase
of a certain beam will cause the lattice to shift in the direction
of the corresponding wave vector.

Since there is no interference for $m_F=0$, by changing the relative
phase $\delta\phi_i=\theta_i-\phi_i$ between corresponding laser
beams of two lattices, we can shift the in-plane lattice and place
its minima between any two out-of-plane lattice minima; consequently, a double-well configuration can be created, as shown in
Fig. (\ref{rep_diagram1}b), where we demonstrate a left-right
double-well configuration. The labels $A-F$ are the six sites in one
unit cell. In Fig. (\ref{rep_diagram1}c), the optical potentials
inside one hexagonal unit cell for the left-right, lower-left, as
well as lower-right double-well configurations are calculated, and
the phase setups are presented, where the double well is formed
by two potential minimums with a lower barrier between them.

The tilt of the double well can be controlled as well through an
adjustment of the relative phases. For example, in the left-right
double well, we can adjust $\delta\phi_1$ while keeping
$\delta\phi_{2,3}=0$, as shown in Fig. (\ref{rep_diagram1}d). The
barrier height of the double well can also be adjusted by changing
the relative laser intensity of in-plane and out-of-plane laser
beams $I_1/I_2$.

\begin{figure*}[tbp]
	\begin{center}
		\includegraphics [width=12cm,height=6.7cm]{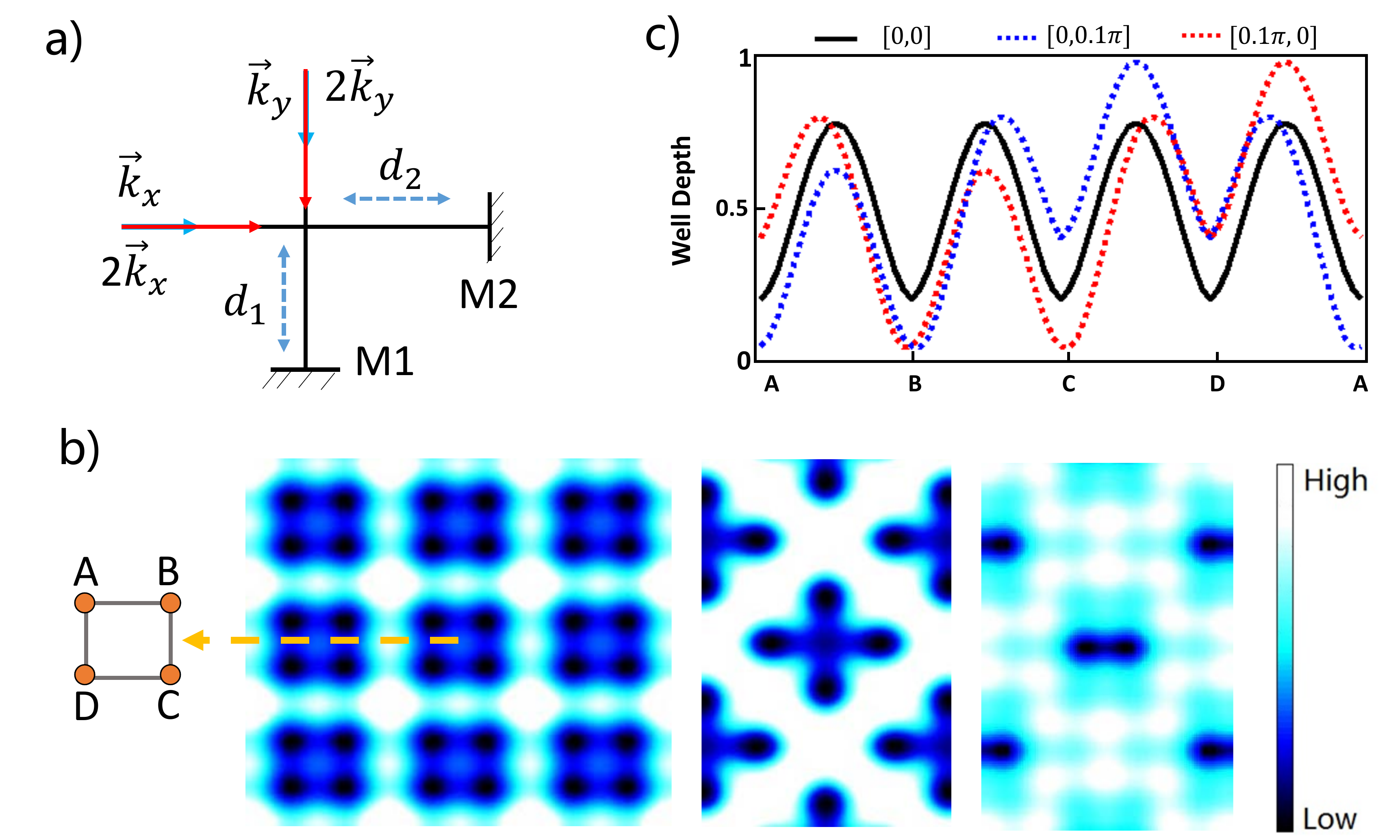}
	\end{center}
	\caption{ (Color online) The isolated square superlattice. (a) The
		setup includes two standing waves for each frequency. Different
		polarizations and detunings can generate different geometries, such
		as (b) the isolated square (left), the isolated crosses (middle) and the isolated double wells (right). The isolated square configuration using in-plane
		polarizations. $\vec{k}_i$ lasers are red detuned, meanwhile
		$2\vec{k}_i$ lasers are blue detuned, with $i=x,y$.
		A unit cell of the superlattice contains four sites $A, B, C$ and $D$.
		(c) Optical potential along the channel $A-B-C-D$ in the isolated square. Adjustment of the relative phases [$\theta_1-\phi_1$, $\theta_2-\phi_2$] changes the relative depth between different sites.
		The x-axis symmetry (red dashed) or the y-axis symmetry (blue dashed) can be broken using different phase setups.
		The black solid line is the optical potential for a symmetric square sublattice.} \label{rep_diagram3}
\end{figure*}

\section{Isolated structures in superlattice}

The construction of isolated structures in 2D requires an overlapping of
multiple commensurate lattices with different periodicities.
In general, the barrier between isolated structures is
created by overlaying the maxima of lattices, while the
structure inside one ``sublattice'' is determined by the interference
pattern of the minima. In the experiment, the tuning of the relative
position between lattices can be realized by changing the relative
phases between laser beams, as demonstrated in the previous section.
Because hexagonal, triangular, and square structures have been
realized in a 2D Bravais lattice, overlaying two such lattices with
different periodicities can generate the isolated hexagonal lattice,
isolated triangular lattice, and isolated square lattice,
respectively.

\subsection{Isolated Square Superlattice}
The 2D square superlattice can be constructed using
eight beams in four perpendicular directions and with two
frequencies $\omega_L$ and $2\omega_L$. Here, we recommend the setup
introduced in Fig. (\ref{rep_diagram3}a) for each frequency, where
the lattice is formed by two standing waves reflected by mirrors
$M1$ or $M2$. For the $x$ direction, the wave vectors of incoming beams
are $\vec{k}_x$ and $2\vec{k}_x$, and the effective distance is
$d_1$ before the beam returns to the atomic cloud. For the $y$
direction, the wave vectors are $\vec{k}_y$ and $2\vec{k}_y$ and the
effective distance is $d_2$, where
$|\vec{k}_y|=|\vec{k}_x|=k={\omega_L}/{c}$, with $c$ representing the speed of light in vacuum. The retro-reflected
beams provide intrinsic topological phase stability while 
providing sufficient freedom for parameter adjustments. Because of the absence
of the cross terms, the total light shifts equal the sum of the
light shifts induced by laser fields with different frequencies.
Therefore, by changing the relative position and light intensity, we
can generate different lattice configurations such as isolated squares, isolated double wells, or isolated crosses, as demonstrated
in Fig. (\ref{rep_diagram3}b).

Here, we focus on the isolated square lattice for the $m_F=0$ atoms.
We can use either in-plane or out-of-plane lattices to generate this
structure. We take an in-plane lattice as an example, where the red-detuned
$\omega_L$ beams $\vec{E}_{1}$ and the blue-detuned $2\omega_L$
beams $\vec{E}_{2}$ are both polarized in the $xy$ plane. The
spatial dependence of the electric field is
\begin{eqnarray}
\begin{split}
\vec{E}_{1}&\propto(e^{i(kx+\theta_1)}+e^{i(-kx+\theta_1+2kd_1)})\hat{y}\\
&+(e^{i(ky+\theta_2+2kd_2)}+e^{i(-ky+\theta_2)})\hat{x}\\
\vec{E}_{2}&\propto(e^{i(2kx+\phi_1)}+e^{i(-2kx+\phi_1+4kd_1)})\hat{y}\\
&+(e^{i(2ky+\phi_2+4kd_2)}+e^{i(-2ky+\phi_2)})\hat{x}\\
\end{split}
\label{doublewellcondition2}
\end{eqnarray}

When $\theta_1=\phi_1,\theta_2=\phi_2$, a symmetric sublattice
structure of isolated squares is formed, as shown in Fig. (\ref{rep_diagram3}b), where the four sites in one unit cell of the
superlattice are labeled $A, B, C,$ and $D$. Notice that the change of
$d_1$ and $d_2$ does not change the geometry but only shifts the
lattice as a whole. The height of the barrier is mostly determined by
the intensity of basis-frequency laser beams, and the depth of the
square lattice wells is primarily affected by the intensity of double-frequency laser beams. The symmetry of the isolated square can be
controlled by shifting the relative position of the two lattices.
For instance, we can break the y-axis symmetry by setting
$\theta_1=\phi_1,\theta_2-\phi_2=0.2\pi$, which makes the site
depths $U_A=U_B\neq U_C=U_D$, or we can break x-axis symmetry by
setting $\theta_1-\phi_1=0.2\pi,\theta_2=\phi_2$, which results in
$U_A=U_D\neq U_C=U_B$. The optical potential along the channel
$A-B-C-D-A$ of the symmetric, x-axis nonsymmetric, and y-axis
nonsymmetric structures is shown in Fig. (\ref{rep_diagram3}c).

\begin{figure*}[tbp]
	\begin{center}
		\includegraphics [width=12cm,height=6.7cm]{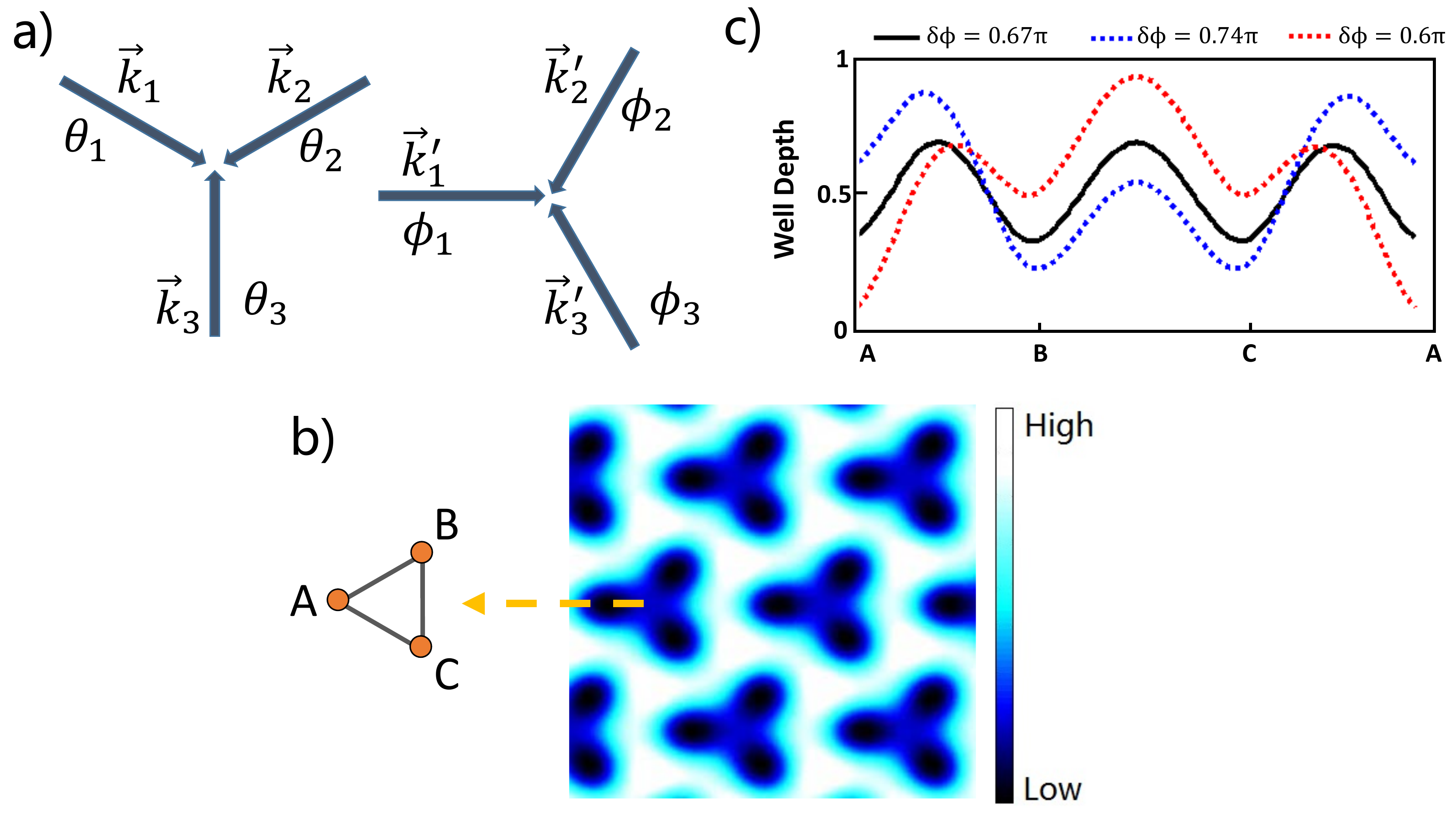}
	\end{center}
	\caption{ (Color online) The isolated triangular superlattice.
		(a) The setup for $m_F=0,\pm 1$ states. The lattice is formed by two sets of three-laser-beam systems. The detuning and polarization depend on the $m_F$ state trapped.
		(b) The 2D optical potential for the isolated triangular configuration. A unit cell of the superlattice contains three sites $A, B$ and $C$.
		(c) Optical potential along the channel $A-B-C-A$. By altering $\delta\phi=\theta_1-\phi_1$, we can make $A$ site depth larger (blue dashed), smaller (red dashed) or equal (black solid) compared to $B$ and $C$ site.} \label{rep_diagram4}
\end{figure*}

\subsection{Isolated Triangular Superlattice}

The setup requires two sets of three-laser-beam systems: one set with
wave vectors $\vec{k_i}$ $(i=1,2,3)$, frequency $\omega_L$, and phases
$\theta_i$, and the other one with $\vec{k'_i}$,
frequency $\omega'_L=\sqrt{3}\omega_L$, and phases $\phi_i$,
as shown in Fig. (\ref{rep_diagram4}a). For the $m_F=\pm1$ state, each
set of beams can be either in-plane or out-of-plane polarized, but
both need to be red detuned. For $m_F=0$ states, each set can be
either out-of-plane polarized and red detuned or in-plane polarized
and blue detuned.

The $\omega_L$ and $\omega'_L$ lasers both generate a triangular
lattice. By changing the relative phases of laser beams and placing the
minimum of the $\omega_L$ lattice at the center of the three
minima of the $\omega'_L$ lattice, we can create the isolated triangular lattice structure, as shown in
Fig. (\ref{rep_diagram4}b), where the three sites in one unit cell
of the superlattice are labeled by $A, B,$ and $C$. The barrier
height inside one triangular ``sublattice'' largely depends on the
intensity of the $\omega'_L$ beams, while the outer barrier
between two triangular ``sublattices'' depends on the intensity of
$\omega_L$ beams.

In Fig. (\ref{rep_diagram4}c), we calculate the 2D optical potential
along the channel $A-B-C-A$, and the lattice potential of a
symmetric isolated triangular is shown with a blue dot-dashed line,
where the sites $A, B,$ and $C$ have the same depth and same barrier
height. By shifting the $\omega_L$ lattice away from
the center, we can break the triangular symmetry and adjust the
relative depth of three minima at will. For example, by
altering $\delta\phi=\theta_1-\phi_1$, we can horizontally shift the
$\omega_L$ lattice and make the depth of site $A$ larger (blue dashed) or
smaller (red dashed) than those of sites $B$ and $C$, which is demonstrated
in Fig. (\ref{rep_diagram4}c). Similarly, a change of
$\theta_2-\phi_2$ or $\theta_3-\phi_3$ can make the depth of site $B$ or $C$ 
different from those of the other two.

\subsection{Isolated Hexagonal Superlattice}

\begin{figure*}[tbp]
	\includegraphics [width=16cm,height=5cm]{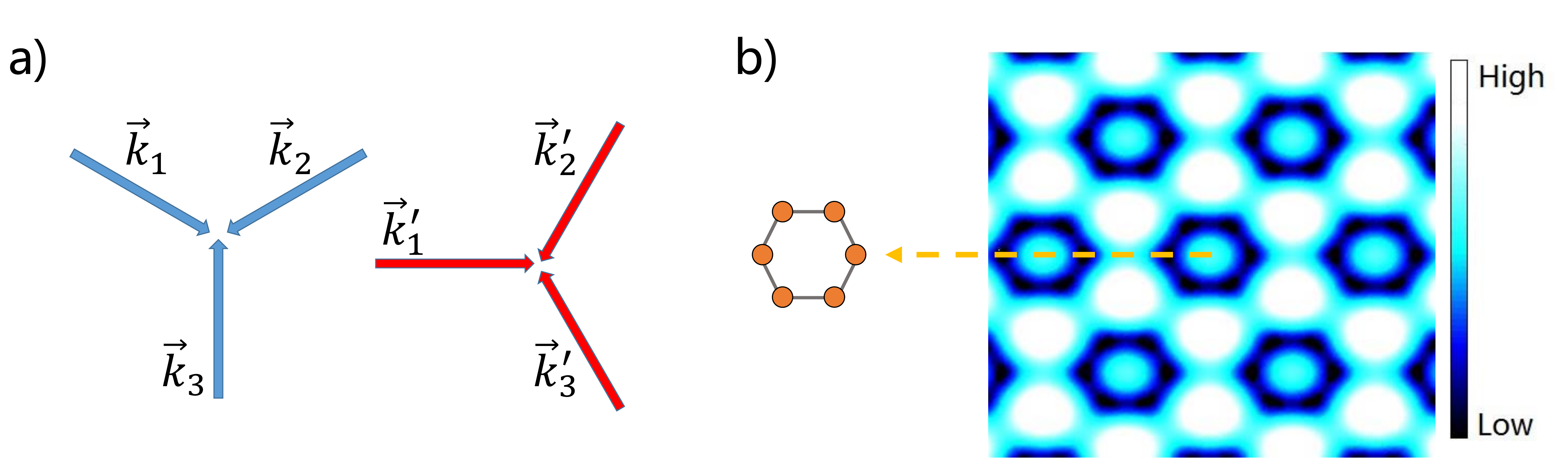}
	\caption{ (Color online) The isolated hexagonal superlattice. (a) The
		setup for $m_F=0,\pm 1$ states. The lattice is formed by two sets of
		three-laser-beam systems, one red detuned and the other blue detuned.
		(b) The 2D optical potential for the isolated hexagonal configuration with a scaled
		colormap.} \label{rep_diagram2}
\end{figure*}

We can use a similar setup to form an optical lattice with isolated
hexagonal structures. The setup requires two sets of
three-laser-beam systems: one set with wave vector $\vec{k_i}$ $(i=1,2,3)$
and frequency $\omega_L$, which is red detuned, and the other one
with $\vec{k'_i}$ and frequency $\omega'_L=\sqrt{3}\omega_L$,
which is taken to be blue detuned. The two sets are perpendicular to
each other, as shown in Fig. (\ref{rep_diagram2}a). For the $m_F=0$
state, all $\omega_L$ and $\omega'_L$ laser beams need to be
out-of-plane polarized, while for $m_F=\pm 1$, each set can be
either in-plane or out-of-plane polarized. The $\omega_L$ and
$\omega'_L$ lasers generate a triangular and hexagonal lattice,
respectively. By changing the relative phase between the two sets of
laser beams and overlapping the maximum of the $\omega'_L$ lattice with
that of the minima of the $\omega_L$ lattice, we can create a 2D lattice with isolated hexagonal structures, as shown in Fig. (\ref{rep_diagram2}b).
\newline
\newline

The essence of creating the isolated structures is to build a high
barrier between ``sublattices,'' which could be a result of
increasing the depth of the ``bigger'' lattice or a result of overlapping the
maxima of different lattices. The pattern inside each
``sublattice,'' however, is a superposition of minima; therefore, a
change of relative depth and position can greatly influence its
shape as well as symmetry. In theory and experiments, the
pattern inside each ``sublattice'' is usually more interesting and
important than the pattern of ``sublattices.'' However, only three
patterns are introduced in this section 
because these three patterns are very basic and representative. Many other structures that are more complex can be created in a similar manner.

\section{Discussions and Conclusions}
We demonstrated the lattice potential for the
ground hyperfine state of $^{87}Rb$ by using the $E^{[2+\epsilon]}$
method to calculate the light shift of hyperfine energy levels and, thereby,
lattice potential. In fact, since the hyperfine corrections are very
small, conventional second-order perturbation theory $E^{[2]}$
applied to hyperfine structures serves the purpose, and we
can replace $\omega_{\rm{F_{ji}}}$ in Eq.(\ref{lightshift1}) with
$\omega_{\rm{J_{ji}}}$. However, when we are concerned with the
different lattice potentials ``felt'' by two hyperfine energy states
of the ground state, only $E^{[2+\epsilon]}$ can provide the difference, 
while $E^{[2]}$ provides identical results for the two states
owing to the absence of hyperfine interactions. Such a situation may
arise when we study the dynamics of a spinor Bose-Einstein condensate (BEC) trapped in an optical
lattice.

The isolated lattices are good candidates for studying multiple-particle entanglement and few-body physics. Once the barrier height
exceeds the chemical potential, the atoms are trapped in one
isolated sublattice, which can be seen as an isolated subsystem, and
its static as well as the dynamic properties can be measured. On one
hand, we can adjust the symmetry of the sublattice and create
different degrees of entanglement between different pairs of atoms.
On the other hand, by loading a BEC into the 2D
isolated ``sublattices,'' we can study the dynamics of this system of coupled
bosonic wells, the pairs and triplets of coupled
wells of which have been studied extensively in the literature \cite{discussion1,discussion2}. This isolated structure
provides unique opportunities to observe many novel phenomena such
as self-trapping, the macroscopic effects of population inversion, and
the Efimov state.

In conclusion, compared with the 1D double-well superlattice studied previously, we proposed new schemes to realize isolated
structures in 2D optical superlattices by overlapping
the maxima of commensurate optical lattices, and various
``sublattice'' patterns are generated by the interference of
minima. By using three different kinds of primitive lattices, we demonstrated
the construction of isolated hexagonal, triangular, and
square superlattices. These configurations not only generate
non-standard lattice geometry, but also provide convenient access for
dynamically tuning the lattice parameters, especially to regulate the symmetry of sublattices by adjusting the relative phase. We believe they give an
opportunity to observe and study novel physics, and in the future,
the lattices presented here can be used for applications in
quantum simulation and quantum information processing, especially for constructing new quantum logic gates.

\begin{small}
	
\end{small}


\begin{thebibliography}{999}
		\bibitem{intro1}
		\href{http://dx.doi.org/10.1103/RevModPhys.86.153} 
		{I. M. Georgescu, S. Ashhab, and F. Nori, Quantum simulation, {\it Rev. Mod. Phys} {\bf 86}, 153 (2004)}
		
		\bibitem{intro2} 
		\href{http://dx.doi.org/10.1103/PhysRevA.70.012306}
		{T. Calarco, U. Dorner, P. S. Julienne, C. J. Williams, and P. Zoller, Quantum computations with atoms in optical lattices: Marker qubits and molecular interactions, {\it Phys. Rev. A} {\bf 70}, 012306 (2004)}
		
		\bibitem{intro3}
		\href{http://dx.doi.org/10.1364/OE.23.010064} 
		{L. Niu, D. Hu, S. Jin, X. Dong, X. Chen, X. Zhou, Excitation of atoms in an optical lattice driven by polychromatic amplitude modulation, {\it Opt. Express} {\bf 23}, 10064 (2015)}
		
		\bibitem{intro4}
		\href{http://dx.doi.org/10.1103/PhysRevA.92.043614} 
		{D. Hu, L. Niu, B. Yang, X. Chen, B. Wu, H. Xiong, and X. Zhou, Long-time nonlinear dynamical evolution for P-band ultracold atoms in an optical lattice, {\it Phys. Rev. A} {\bf 92} 043614 (2015)}	
		
		\bibitem{introSFMI} 
		\href{http://dx.doi.org/10.1038/415039a}
		{M. Greiner, O. Mandel, T. Esslinger, T. W. Hansch, and I. Bloch, Quantum phase transition from a superfluid to a Mott insulator in a gas of ultracold atoms, {\it Nature} {\bf 415} 39 (2002)}
		
		\bibitem{introtriangular}
		\href{http://dx.doi.org/10.1088/1367-2630/12/6/065025}
		{C. Becker, P. Soltan-Panahi, J. Kronj\"ager, S. D\"orscher, K. Bongs and K. Sengstock, Ultracold quantum gases in triangular optical lattices, {\it New J. Phys.} {\bf 12} 065025 (2010)}
		
		\bibitem{introdoublewell} 
		\href{http://dx.doi.org/10.1103/PhysRevA.73.033605}
		{J. Sebby-Strabley, M. Anderlini, P. S. Jessen, and J. V. Portol, Lattice of double wells for manipulating pairs of cold atoms, {\it Phys. Rev. A} {\bf 73} 033605 (2006)}
		
		
		\bibitem{introkagome} 
		\href{http://dx.doi.org/10.1103/PhysRevLett.93.030601}
		{L. Santos, M. A. Baranov, J. I. Cirac, H. U. Everts, H. Fehrmann, and M. Lewenstein, Atomic quantum gases in Kagom{\'e} lattices, {\it Phys. Rev. Lett.} {\bf 93} 030601 (2004)}
		
		\bibitem{introkagome1} 
		\href{http://dx.doi.org/10.1103/PhysRevLett.108.045305}
		{G. B. Jo, J. Guzman, C. K. Thomas, P. Hosur1, A. Vishwanath, and D. M. Stamper-Kurn, Ultracold atoms in a tunable optical kagome lattice, {\it Phys. Rev. Lett.} {\bf 108} 045305 (2012)}
		
		\bibitem{quanOneD} 
		\href{http://dx.doi.org/10.1103/PhysRevLett.91.107902}
		{J. K. Pachos and P. L. Knight, Quantum computation with a one-dimensional optical lattice, {\it Phys. Rev. Lett.} {\bf 91} 108103 (2003)}
		
		\bibitem{quanlogic} 
		\href{http://dx.doi.org/10.1103/PhysRevLett.82.1060}
		{G. K. Brennen, C. M. Caves, P. S. Jessen, and I. H. Deutsch, Quantum logic gates in optical lattices, {\it Phys. Rev. Lett.} {\bf 82} 1060 (1999)}
		
		\bibitem{squre} 
		\href{http://dx.doi.org/10.1103/PhysRevA.79.022309}
		{L. Jiang, A. M. Rey, O. Romero-Isart, J. J. Garca-Ripoll, A. Sanpera, and M. D. Lukin, Preparation of decoherence-free cluster states with optical superlattices, {\it Phys. Rev. A} {\bf 79.2} 022309 (2009)}
		
		\bibitem{isolateTriangular1} 
		\href{http://dx.doi.org/10.1103/PhysRevA.63.013604}
		{K. Nemoto, C. A. Holmes, G. J. Milburn, and W. J. Munro, Quantum dynamics of three coupled atomic Bose-Einstein condensates, {\it Phys. Rev. A} {\bf 63} 013604 (2000)}
		
		\bibitem{isolateTriangular2} 
		\href{http://dx.doi.org/10.1103/PhysRevA.73.061604}
		{M. Hiller, T. Kottos, and T. Geisel, Complexity in parametric Bose-Hubbard Hamiltonians and structural analysis of eigenstates, {\it Phys. Rev. A} {\bf 73} 061604(R) (2006)}
		
		\bibitem{isolateTriangular3} 
		\href{http://dx.doi.org/10.1103/PhysRevLett.99.020401}
		{A. R. Kolovsky, Semiclassical quantization of the Bogoliubov spectrum, {\it Phys. Rev. Lett.} {\bf 99} 020401 (2007)}
		
		\bibitem{isolateTriangular4} 
		\href{http://dx.doi.org/10.1103/PhysRevA.65.013601}
		{R. Franzosi and V. Penna, Self-trapping mechanisms in the dynamics of three coupled Bose-Einstein condensates, {\it Phys. Rev. A} {\bf 65} 013601 (2001)}
		
		\bibitem{isolateTriangular5} 
		\href{http://dx.doi.org/10.1038/nphys3211}
		{P. Hsieh, C. Chung, J. McMillan, M. Tsai, M. Lu, N. Panoiu, and C. W. Wong, Photon transport enhanced by transverse Anderson localization in disordered superlattices, {\it Nat. Phys.} {\bf 11} 268 (2015)}
		
		\bibitem{efimov} 
		\href{http://dx.doi.org/10.1126/science.aaa5601}
		{M. Kunitski, S. Zeller, J. Voigtsberger, A. Kalinin, L. P. H. Schmidt, M. Sch{\"o}ffler, A. Czasch, W. Sch{\"o}llkopf, R. E. Grisenti, T. Jahnke, et al., Observation of the Efimov state of the helium trimer, {\it Science} {\bf 348} 551 (2015)}
		
		\bibitem{magic}	
		\href{http://dx.doi.org/10.1103/PhysRevA.81.012115}
		{X. Zhou, X. Xu, X. Chen, J. Chen, Magic wavelengths for terahertz clock transitions, {\it Phys. Rev. A} {\bf 81} 012115 (2010)}
		
		\bibitem{Esecondorder} 
		\href{http://dx.doi.org/10.1016/j.physleta.2015.03.024}
		{X. Xu, B. Qing, X. Z. Chen, X. J. Zhou, A simplified method for calculating the ac Stark shift of hyperfine levels of alkali-metal atoms, {\it Phys. Lett. A} {\bf 379} 1347 (2015)}
		
		\newpage
		
		\bibitem{zeeman} 
		\href{http://dx.doi.org/10.1038/nphys1916}
		{P. Soltan-Panahi, J. Struck, P. Hauke, A. Bick, W. Plenkers, G. Meineke, C. Becker, P. Windpassinger, M. Lewenstein and K. Sengstock, Multi-component quantum gases in spin-dependent hexagonal lattices, {\it Nat. Phys.} {\bf 7} 434 (2011)}
		
		\bibitem{discussion1} 
		\href{http://dx.doi.org/10.1016/S0370-1573(98)00014-3}
		{A. S. Parkins and D. F. Walls, The physics of trapped dilute-gas Bose–Einstein condensates, {\it Phys. Rep.} {\bf 303} 1 (1998)}
		
		\bibitem{discussion2} 
		\href{http://dx.doi.org/10.1103/RevModPhys.71.463}
		{F. Dalfovo, S. Giorgini, L. P. Pitaevskii, and S. Stringari, Theory of Bose-Einstein condensation in trapped gases, {\it Rev. Mod. Phys.} {\bf 71} 463 (1999)}
	\end{thebibliography}
\end{document}